# Design and Fabrication of PERC-Like CdTe Solar Cells Using Micropatterned $Al_2O_3$ Layer

Etee Kawna Roy[1], Kaden Powell[1], Chungho Lee[2], Gang Xiong[2], and Heayoung Yoon[1]

[1]Department of Electrical and Computer Engineering, University of Utah, Salt Lake City, UT 84112, USA

[2]California Technology Center, First Solar Inc., Santa Clara, CA, USA.

*Abstract* —Recent studies have investigated novel strategies to further improve the limited $V_{oc}$ of CdTe solar cells via increased carrier lifetime and doping density of CdTe thin films. Among various metal oxides, aluminum oxide ($Al_2O_3$) is a promising passivation candidate, where the negatively charged $Al_2O_3$ layer repels the minority carrier in CdTe and $Al_2O_3$ provides a chemically passivating interface, increasing the carrier lifetime. Despite the continuing efforts, an optimized back-contact architecture to improve the $V_{oc}$ while maintaining high $J_{sc}$ and *FF* is still under development. In this work, we report the design, fabrication, and characterization of PERC-like CdTe solar cells, where an $Al_2O_3$ passivation layer is patterned using laser-beam lithography. Our process enables reproducible patterning on a rough surface CdTe while maintaining the size of the array in the design. Analysis of CdTe PERC devices (As-doped) shows a notably different $V_{oc}$ trend compared to *FF* and $J_{sc}$, independent of the patterned array structures used in this study. The subsurface electronic structure of CdTe and the interplay between carrier selectivity and collection of the patterned $Al_2O_3$ could be responsible for the observed PV characteristics.

## I. INTRODUCTION

Thin-film CdTe solar cells are a competitive photovoltaic (PV) technology that can meet the rapidly growing societal demand for energy owing to cost-effective manufacturing and relatively short energy payback time [1, 2]. At a record efficiency of 22.1 %, significant attention has been devoted to further improving the power conversion efficiencies of CdTe-based solar cells [3-5]. An optimized front contact that leads the $J_{sc}$ over 31 mA/cm$^2$ and a fill factor (FF) above 79 % were reported [6, 7]. An open-circuit voltage ($V_{oc}$) over 1 V was achieved with As-doped CdTe single crystals. Despite the continuing efforts, high $V_{oc}$ (> 1 V) has not yet been observed in thin-film polycrystalline CdTe PVs. The underlying physical mechanisms responsible for the $V_{oc}$ loss are not presently well understood.

Recent studies have demonstrated novel device architectures and passivation strategies to enhance the $V_{oc}$ via increasing carrier lifetime and doping density of CdTe thin films. Among various metal oxides, $Al_2O_3$ is a promising passivation candidate, where the negatively charged $Al_2O_3$ layer (fixed charge density of $10^{12} \sim 10^{13}$ cm$^{-2}$) could repel the minority carrier (electrons) in CdTe, increasing the carrier lifetime [7, 8]. Previous studies also suggested that $Al_2O_3$ could passivate the CdTe surface via chemical reactions during the fabrication processes. This configuration is similar to a passivated emitter and rear contact (PERC) design, frequently used in Si photovoltaic (PV) technology [9, 10]. An improved $V_{oc}$ was reported with a conformal $Al_2O_3$ (< 5 nm) on CdTe, yet the $J_{sc}$ and FF suffer from the tunneling barrier [11-18]. On the other hand, Kephart and co-workers used a 20 nm-thick $Al_2O_3$ with micropatterning. A remarkably high lifetime ($\tau_2$ > 400 ns) was measured for double heterostructures, but no consistent improvement of $V_{oc}$ was observed for the Cu-doped CdTe solar cells [11].

This work reports the design and fabrication of PERC-like CdTe solar cells (CdTe PERC), where the $Al_2O_3$ passivation layer is patterned via laser-beam lithography. We optimize the beam dose to polymerize the photoresist on a rough surface CdTe while maintaining the size of the array in the design. We measure the dark and light *I-V*s of As-doped CdSeTe PERC having different hole diameters and pitches (i.e., the distance between adjacent holes). Quantitative analysis shows the impact of $Al_2O_3$ patterns on $V_{oc}$ is notably different from other device parameters of $J_0$, $J_{sc}$, and *FF*. We discuss a similar $V_{oc}$ trend observed after annealing (70°C for 2.5 hours) of the complete CdTe PERC. Our results indicate the electrically-active $Al_2O_3$ patterns, in conjunction with the defect chemistry in CdTe bulk, could play an essential role in determining the $V_{oc}$ of advanced CdTe-based solar cells.

## II. EXPERIMENT

Figure 1(a) displays a schematic of a CdTe PERC device consisting of a thin-film CdTe solar cell and a patterned $Al_2O_3$. The As-doped CdSeTe/CdTe absorber layer was synthesized on a stack of buffer/TCO [transparent conductive oxide]/glass substrate. A scanning electron microscopy (SEM) image shows a representative $Al_2O_3$ hole array on CdTe fabricated in this work (Figure 1b). The same-size individual holes are uniformly distributed in a square pattern across the entire device. This study used three different hole sizes (10 μm, 20 μm, and 40 μm). The distance between the adjacent holes ranges from 10 μm to 320 μm, introducing different sizes of the exposed CdTe area to metal contact.

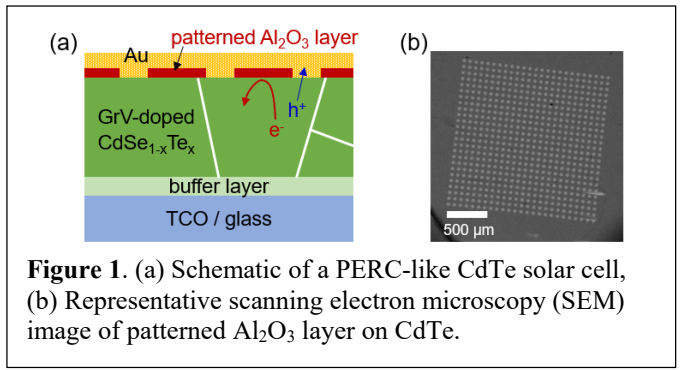

**Figure 1**. (a) Schematic of a PERC-like CdTe solar cell, (b) Representative scanning electron microscopy (SEM) image of patterned $Al_2O_3$ layer on CdTe.

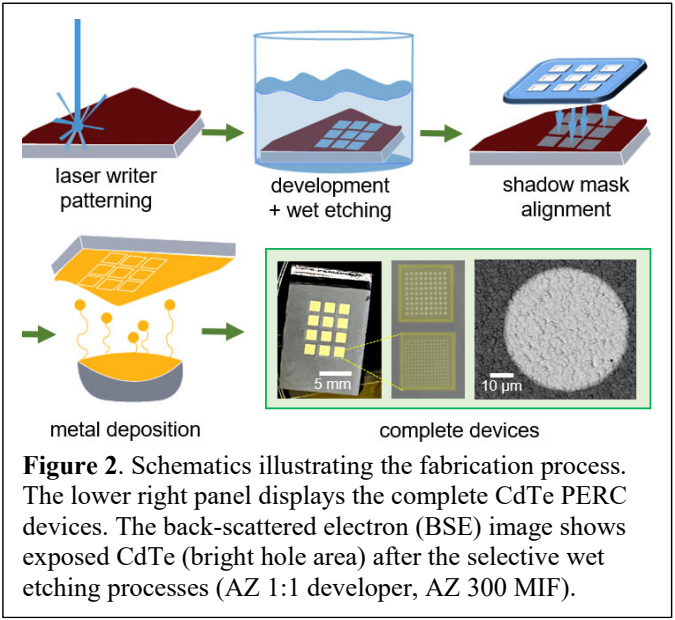

**Figure 2.** Schematics illustrating the fabrication process. The lower right panel displays the complete CdTe PERC devices. The back-scattered electron (BSE) image shows exposed CdTe (bright hole area) after the selective wet etching processes (AZ 1:1 developer, AZ 300 MIF).

Figure 2 illustrates the fabrication procedures to produce patterned $Al_2O_3$ back contacts. As-doped CdSeTe/CdTe solar cells were obtained from First Solar that was coated with a 20 nm $Al_2O_3$ layer. The samples were cleaned with acetone and isopropyl alcohol (IPA), and blown dry with nitrogen ($N_2$). The samples were baked on a hot plate at 100°C for 60 seconds and cooled down before spin coating of photoresist. A thin layer of positive photoresist (Shipley1813) was coated on the sample at a spin speed of 3,000 rpm (revolution per minute) for 60 seconds, followed by soft baking for another 60 seconds on the hot plate at 100°C. A laser writer (Heidelberg μPG 101) was used to pattern the hole array design (L-Edit) on the photoresist-coated $Al_2O_3$ on CdTe.

We performed a series of control experiments to optimize the laser dose that could polymerize the photoresist on a rough surface CdTe while maintaining the size of the array of the design [19]. We found a relatively high beam energy (an effective dose of 13.5 mW) is needed for CdTe compared to traditional Si flat samples (10 mW). The photoresist was developed in a solution (AZ Developer 1:1) for 60 seconds and soaked in DI water. We used another solution (AZ 300 MIF) to selectively etch the exposed $Al_2O_3$ on CdTe after the photoresist development. This chemical contains a small amount of tetramethyl ammonium hydroxide (< 3 % TMAH), which has a faster etch rate for $Al_2O_3$ than photoresist. The $Al_2O_3$ etching was conducted for 20 minutes, followed by the photoresist removal and cleaning using acetone, IPA, deionized water, and $N_2$-blown dried.

The metal contact for each device was designed to be 2 mm × 2 mm in size. We fabricated a shadow mask using a stainless-steel foil (40 μm) containing a square array (3 × 4) with a spacing of 1 mm. This shadow mask was aligned to the hole array patterns on $Al_2O_3$/CdTe (top right in Figure 2). The sample was placed on a holder in an electron beam evaporator. Thin films of Cu/Au (3 nm/80 nm) were deposited on the patterned $Al_2O_3$/CdTe through the shadow mask, which served as a back contact. A front indium contact was formed on TCO after removing the CdTe segment using a razor blade. Figure 2 displays the complete CdTe PERC devices.

The dark and light current-voltage characteristics were measured using a probe station (dark enclosure) connected to a semiconductor analyzer (Agilent 4145C). Each probe tip (25 μm diameter W-tip) was placed on top and bottom metal contact while recording the $I$-$V$s using the LabVIEW program. Before the measurements, the solar simulator (G2V) was calibrated to 1-sun using a reference cell (Newport; 91150V).

## III. RESULTS AND DISCUSSION

We analyze the dark and light $I$-$V$s measured from 36 CdTe PERC devices having various hole array patterns. All devices consisted of a patterned area in the center (1 mm × 1 mm) of the metal contact (2 mm × 2 mm). For a statistical comparison, we estimate the exposed CdTe area of each device from the design, where the Cu/Au metal is directly deposited on CdTe without the $Al_2O_3$ layer. Figure 3 summarizes the parameters ($I_0$, $I_{sc}$, $FF$, and $V_{oc}$) of CdTe PERC.

The distribution of leakage current of our devices (Figure 3a) is relatively constant, supporting the fidelity of the fabrication processes developed in this work. The magnitudes of $I_{sc}$ and $FF$ proportionally increase with the exposed CdTe area. As seen in Figure 3(b), the $I_{sc}$ of the CdTe PERC having an exposed CdTe area of 1 % is 1 µA. This current increases to 10 mA when the $Al_2O_3$ layer is fully removed in the patterned area (1 mm × 1 mm). The *FF* increases to a factor of two (approximately 30 % to 60 %) with the $Al_2O_3$ removal. These trends are expected as the electrically resistive $Al_2O_3$ layer (20 nm thick) can interfere with the photocarrier collection (hole carriers) from CdTe to metal contact.

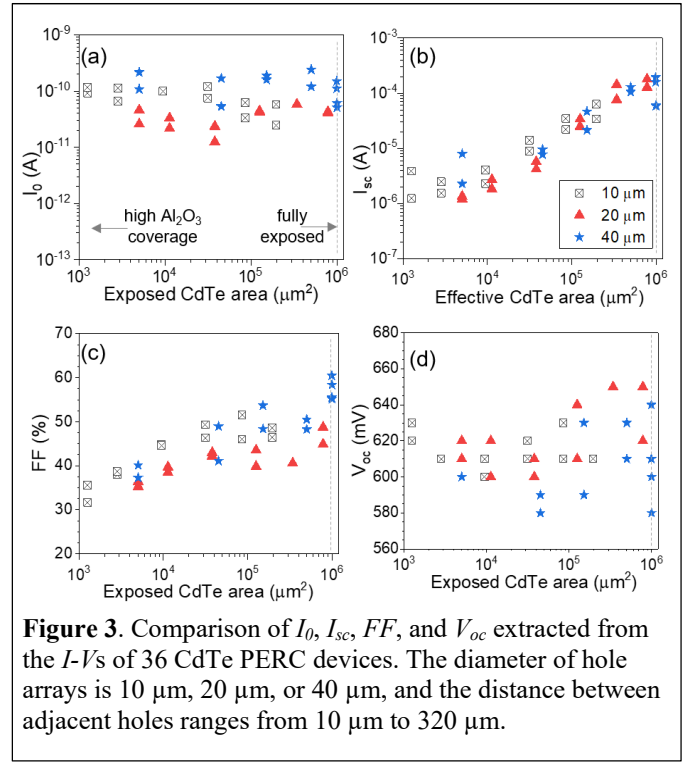

**Figure 3.** Comparison of $I_0$, $I_{sc}$, $FF$, and $V_{oc}$ extracted from the $I$-$V$s of 36 CdTe PERC devices. The diameter of hole arrays is 10 μm, 20 μm, or 40 μm, and the distance between adjacent holes ranges from 10 μm to 320 μm.

Interestingly, the $V_{oc}$ trend (Figure 3d) for the As-doped CdTe PERC is notably different from that of $I_{sc}$ and *FF*. The $V_{oc}$ ranges from 580 mV to 660 mV, which seems independent to the exposed CdTe areas. In a close look, the $V_{oc}$ slightly decreases from ≈ 640 mV to ≈ 590 mV when the CdTe area increases from $10^3$ µm$^2$ (1 % of the total device area) to $10^4$ µm$^2$ (10 %). The $V_{oc}$ increases as high as 660 mV with the increase of the exposed CdTe area (40 % ~ 80 %). The $V_{oc}$ of the control devices, which have the fully open CdTe area (100 %), ranges from 580 mV to 640 mV, still below the highest $V_{oc}$ observed with the patterned $Al_2O_3$ PERC devices. We note that the $V_{oc}$ trend of As-doped PERC is also significantly different from our Cu-doped CdTe PERC, where the $V_{oc}$ decreases with the increased CdTe exposed area (not shown in the data). Our results indicate the electrically-active $Al_2O_3$ patterns could play an important role in determining the $V_{oc}$ of advanced CdTe-based solar cells.

We have examined the performance of the As-doped CdTe PERC solar cells under heating. We annealed the complete PERC devices at 70 °C for 2.5 hours in this preliminary experiment. The diameter of the patterned $Al_2O_3$ is 10 µm, and the adjacent hole distance ranges from 10 µm to 320 µm. Figures 4 (a, b) shows the light *I-V*s collected "before" and "after" the heating. Overall, the PERC devices preserve good diode behaviors after the heating.

Figures 4 (c ~ f) compare the device parameters of the saturation current ($I_s$), $I_{sc}$, *FF*, and $V_{oc}$ extracted from the *I-V*s. The magnitudes of the $I_s$ and $V_{oc}$ are relatively constant, while the $I_{sc}$ and *FF* increase with the exposed CdTe area, with similar trends observed in Figure 3. The magnitudes of $I_s$, $I_{sc}$, and *FF* show only slight variation after the heating at 70 °C. In contrast, the V $_{oc}$ of the CdTe PERC increases after annealing with a magnitude as high as ≈ 10 % of its initial value (e.g., 600 mV to 650 mV). Presumably, the group-V dopant (As) in the CdTe bulk could be more activated, or the defects near $Al_2O_3$ may be further passivated during the annealing [20]. Further studies are in progress to gain a better understanding of such heating effects and back-contact passivation.

## IV. CONCLUSIONS

In summary, we have demonstrated the fabrication of CdTe PERC solar cells with a patterned $Al_2O_3$ passivation layer. Our process based on laser-beam lithography enables reproducible patterning on a rough surface CdTe while maintaining the size of the array in the design. Quantitative analysis of the As-doped CdTe PERC shows a proportional increase of *FF* and $J_{sc}$ with an exposed CdTe area. We have observed that the complete CdTe PERC annealing increases the $V_{oc}$ by ≈ 10 % of their initial magnitudes, whereas the $I_s$, $I_{sc}$, and *FF* show negligible changes. Our results indicate that the subsurface electronic properties of CdTe and the interplay between carrier selectivity and collection of the patterned $Al_2O_3$ could be responsible for the observed PV characteristics.

ACKNOWLEDGEMENT

The authors thank B. Baker, A. Hurlbut, P. Perez, D. Albin, A. Chowdhury, and D. Magginetti for experimental assistance and valuable discussions in this work. This research was supported by the U.S. Department of Energy's Office of Energy Efficiency and Renewable Energy (EERE) under the DE-FOA-0002064 program award number DE-EE0008983. We acknowledge support in part by the National Science Foundation (NSF) CAREER Award No. 2048152.

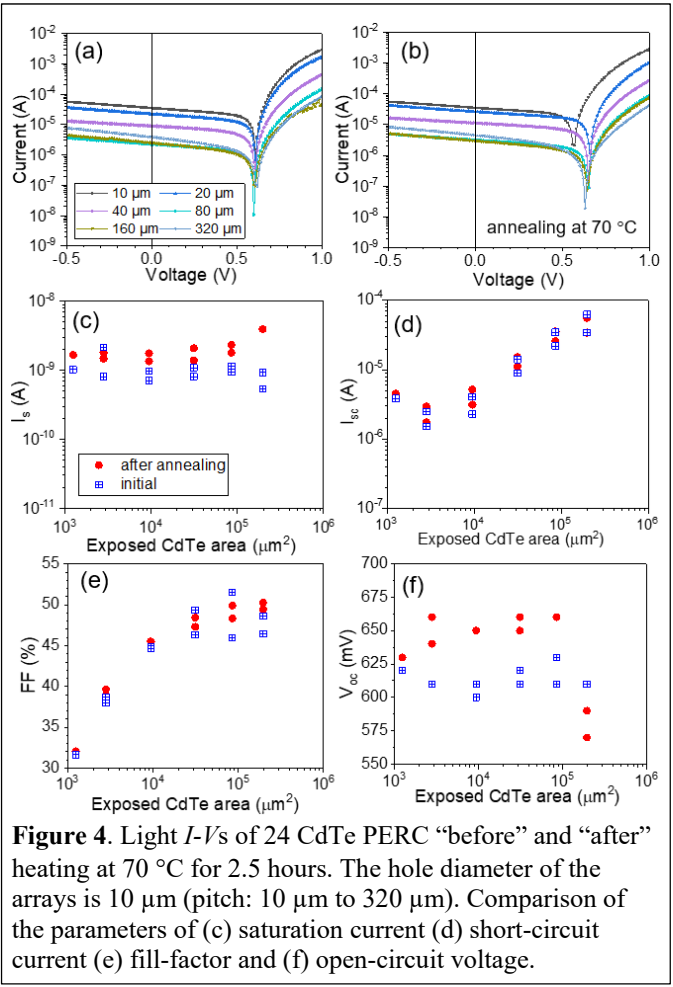

**Figure 4**. Light *I-V*s of 24 CdTe PERC "before" and "after" heating at 70 °C for 2.5 hours. The hole diameter of the arrays is 10 µm (pitch: 10 µm to 320 µm). Comparison of the parameters of (c) saturation current (d) short-circuit current (e) fill-factor and (f) open-circuit voltage.